\documentclass[aps,pre,showpacs,groupedaddress,nofootinbib,twocolumn]{revtex4}
\usepackage{graphicx,amsmath,amssymb}
\def\be{\begin{equation}}
\def\ee{\end{equation}}
\def\bea{\begin{eqnarray}}
\def\eea{\end{eqnarray}}
\def\ba{\begin{array}}
\def\ea{\end{array}}
\def\<{\left\langle}
\def\>{\right\rangle}
\def\({\left(}
\def\){\right)}
\def\e{{\rm e}}

\begin{document}
\title{
Accuracy and range of validity of the Wigner surmise for mixed symmetry classes\\
in random matrix theory
}
\author{Shinsuke M. Nishigaki}
\affiliation{
Graduate School of Science and Engineering,
Shimane University,
Matsue 690-8504, Japan
}
\date{September 4, 2012; Revised: December 3, 2012}
\begin{abstract}
Schierenberg {\it et al.} [Phys.~Rev.~E {\bf 85}, 061130 (2012)] 
recently applied the Wigner surmise,
i.e., substitution of $\infty\times\infty$ matrices by their $2\times 2$ counterparts
for the computation of level spacing distributions,
to random matrix ensembles in transition between two universality classes.
I examine the accuracy and the range of validity of the surmise for the crossover
between the Gaussian orthogonal and unitary ensembles 
by contrasting them with the large-$N$ results that I evaluated using
the Nystr\"{o}m-type method for the Fredholm determinant.
The surmised expression at the best-fitting parameter
provides a good approximation for $0\lesssim s\lesssim 2$, 
i.e., the validity range of the original surmise.
\end{abstract}
\pacs{
02.50.-r, 
05.45.Mt  
}
\maketitle
 
Recently, Schierenberg {\it et al.}~\cite{sbw} publicized an exhaustive study on
the application of the Wigner surmise \cite{wig} 
to the computation of
level spacing distributions (LSDs) of random matrix ensembles in ``mixed symmetry classes,''
i.e.,
\be
H=H_\beta+ \lambda H_{\beta'}\ \in \left\{2\times2\ \mbox{Hermitian matrices}\right\}
\label{WS}
\ee
with $H_\beta$ and $H_{\beta'}$ taken from the Gaussian measures $dH_\beta\,\e^{-(1/2){\rm tr}H_\beta^2}$
of Dyson indices $\beta$ and $\beta'$, respectively,
succeeding the works by Lenz and Haake and others \cite{lh,abps}.
Although the substitution of random matrices of infinite rank
with those of the smallest possible rank appears too bold
an ansatz, the resultant distributions
$P(s)$ of unfolded level spacings $s=(\epsilon_{i+1}-\epsilon_i)/\Delta$ 
[$\epsilon_i\in{\rm Spec}(H_\beta)$, $\Delta$: mean level spacing around $\epsilon_i$]
for the three Dyson universality classes $\beta=1,2$ and 4 
are well known to be in good agreement with the
exact large-$N$ results \cite{jmms,tw},
especially in the range $0\lesssim s\lesssim 2$ (Fig.~1).
\begin{figure}[b]
\includegraphics{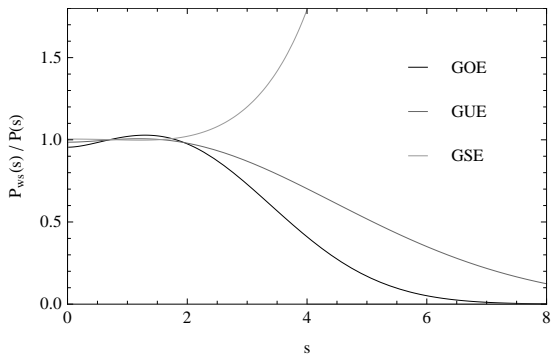}
\caption{
Ratios between Wigner-surmised LSDs $P_{\rm ws}(s)$ and 
their exact forms $P(s)$ for GOE, GUE, and GSE ($\beta=1, 2, 4$).
}
\end{figure}

The intuitive reason for this is that the small-$s$ behaviors of
LSDs are dominated by the
repulsion of two (otherwise colliding) eigenvalues of a Hamiltonian under perturbation.
Eigenvalues adjacent to the two anticrossing eigenvalues are
expected to have very little influence on this universal repulsive behavior,
leading to the validity of the $2\times 2$ approximation for relatively small spacing $s$.
However, even this intuitive account can hardly be taken for granted
in the case of Hamiltonians in universality crossover.
Nevertheless, the authors of \cite{sbw} have reported that
the numerically generated spectrum of random matrices of large rank $N (\sim 400)$,
\be
H=H_\beta+ \alpha H_{\beta'}\ \in \left\{N\times N\ \mbox{Hermitian matrices}\right\}~,
\label{N}
\ee
can be fitted well to the Wigner-surmised expression from (\ref{WS}) whose $\lambda$
parameter is close to 
$\Lambda\equiv (\bar{s}_\beta/\Delta) \alpha$
(Figs.~5, 7, 9, and 10 of \cite{sbw}).
Here, $\bar{s}_\beta$ is the mean level spacing of the unperturbed $2\times 2$ random matrix $H_\beta$.

On the other hand, I have recently evaluated LSDs 
for 
crossovers between Gaussian orthogonal (GOE) and unitary
(GUE) ensembles and between Gaussian symplectic ensembles (GSE) and GUEs
($\beta=1,4,\ \beta'=2$) in the large-$N$ limit
by applying the Nystr\"{o}m-type method \cite{bor} to the Fredholm determinant
$E(s)={\rm Det} (I-\hat{K}_s)$ 
of the dynamical sine kernel \cite{mp}, $(\hat{K}_s\cdot f)(x) \equiv \int_0^s K(x,y) f(y) dy$.
Although this method of discretizing the Fredholm determinant 
by the Gaussian quadrature using $m$ Legendre points $\{x_i\}$
and corresponding weights $\{w_i\}$,
\be
{\rm Det}(I-\hat{K}_s)\simeq \det \left[\delta_{ij}-K(x_i,x_j) \sqrt{w_i\,w_j}\right]_{i,j=1}^m~,
\label{Nystrom}
\ee
is also a numerical approximation, uniform convergence 
to the exact LSD is proven to hold as the number $m$ of quadrature points increases.
The convergence is exponentially fast, i.e., the approximation error decays as $O(\e^{-{\rm const.}m})$.
It is sufficient to take $m$ to be $100\sim200$, as
the relative shifts of $E(s)$ (for the GOE-GUE crossover) 
under the increment of $m$ from 100 to 200 are as small as
$10^{-8\sim 10}$ for $E(1)$,
$10^{-7\sim 8}$ for $E(2)$,
$10^{-6\sim 7}$ for $E(3)$, and
$10^{-4\sim 5}$ for $E(4)$.
This stability
ensures that numeric values obtained for $P(s)$ by the Nystr\"{o}m-type method at $m=200$ (used below) 
can practically be regarded as exact for fitting to the actual spectra of physical systems.
In view of this recent progress, it is now possible to 
examine the accuracies of Wigner-surmised expressions reported in \cite{sbw}
and their range of validity.
In this Brief Report, I 
\begin{widetext}

\begin{figure}[t]
\hspace{6mm}
\includegraphics{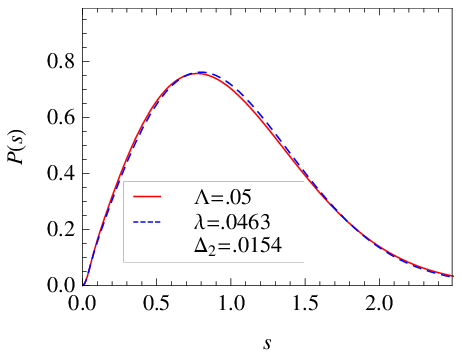}\hspace{-2.7mm}
\includegraphics{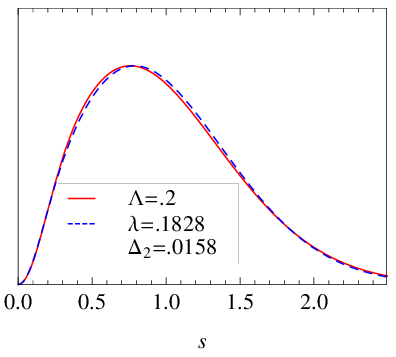}\hspace{-9mm}
\includegraphics{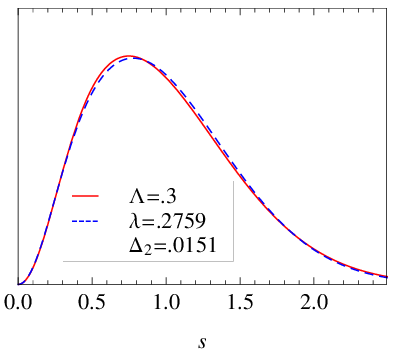}\hspace{-9mm}
\includegraphics{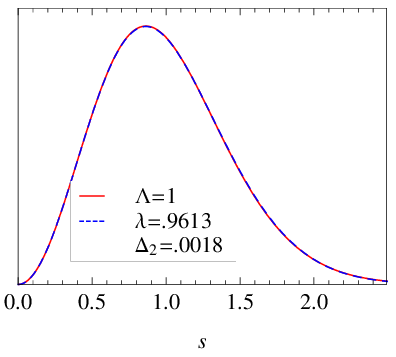}\hspace{-9mm}\\
\hspace{-5mm}
\includegraphics{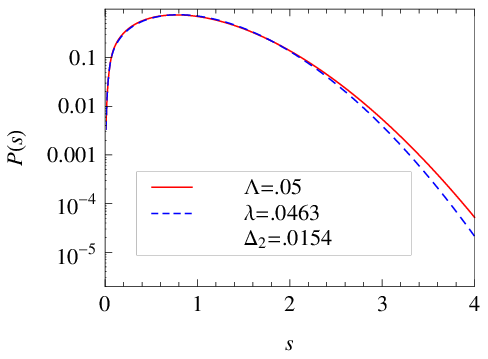}\hspace{-1mm}
\includegraphics{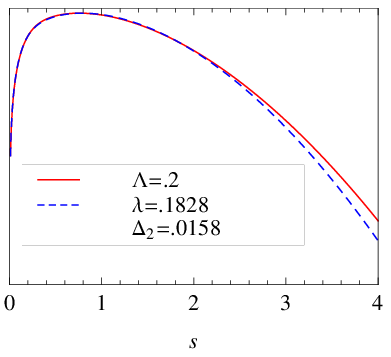}\hspace{-9mm}
\includegraphics{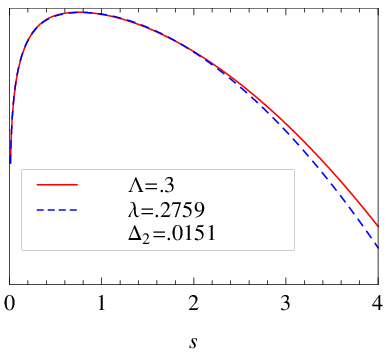}\hspace{-9mm}
\includegraphics{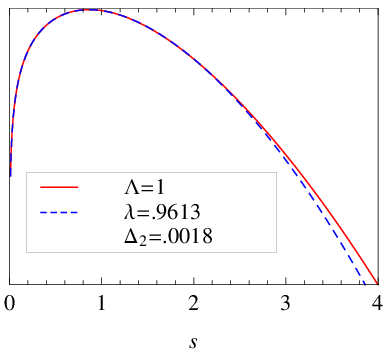}\hspace{-9mm}
\caption{(Color onine) 
Linear and log plots of LSDs $P(s)$ for GOE-GUE crossover evaluated by the Nystr\"{o}m-type method at $m=200$ (red liness)
\cite{nis}, and
Wigner-surmised forms $P_{\rm ws}(s)$ (blue, dotted lines) at optimally fitting values of $\lambda$.
$\Delta_2$ denotes the $L^2$ distance between $P(s)$ and $P_{\rm ws}(s)$.
}
\end{figure}
\end{widetext}

\begin{figure}[hb]
\includegraphics{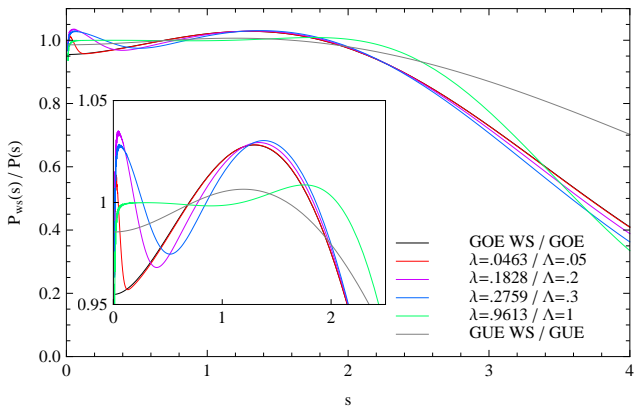}
\caption{(Color onine) Ratios $P_{\rm ws}(s)/P(s)$ for GOE-GUE crossover, including the two limiting cases GOE and GUE.}
\end{figure}
\noindent
shall concentrate on the simplest case of transition between two symmetry classes, namely,
LSD for the GOE-GUE crossover [(2.33)--(2.35) of \cite{sbw}], as an example.

Note that the crossover parameter $\Lambda$ used in \cite{sbw} (with $\bar{s}_1=\sqrt{\pi}$) 
is equal to $\sqrt{2\pi}$ times the $\rho$ parameter used in \cite{nis,meh}.
In Fig.~2, I plot
LSDs $P(s)$ at the choice of \cite{sbw}, 
$\Lambda=0.05$, 0.1, 0.2, and 1,
together with respective Wigner-surmised LSDs $P_{\rm ws}(s)$ at 
$\lambda=0.0463$, 0.1828, 0.2759, and 0.9613 that minimize the $L^2$ distances
between $P(s)$ and $P_{\rm ws}(s)$.
These $\lambda$ and $\Delta_2$ values are consistent with those 
obtained from fitting to the spectra of random $400\times400$ matrices
($\lambda=0.047$, 0.182, 0.276, 0.959)
shown in Fig.7 (top) of \cite{sbw}, 
which should be substituted with Fig.~2 of this paper.
The ratios $P_{\rm ws}(s)/P(s)$ are shown in Fig.~3.\\
\indent
In the range $0\lesssim s\lesssim 2$, i.e., the validity range of the original Wigner surmise,
the best-fitting curve $P_{\rm ws}(s)$ for each $\lambda$
is hardly discernible from $P(s)$ at corresponding $\Lambda$ and serves as its
good approximation,
and the agreement deteriorates for $s\gtrsim 2$.
Consequently, when one applies Wigner-surmised LSDs to, e.g., the Dirac spectra of QCD-like theories
at imaginary baryon chemical potential \cite{bbms}, 
the region $s\gtrsim 2$ should be avoided for fitting.
And even if one restricts the fitting region to $0\lesssim s\lesssim 2$,
an obvious shortcoming of the Wigner surmise applied to the crossover ensembles
lies in the difference between $\Lambda$ and $\lambda$: 
if one tries to determine the values of $\alpha$ or $\Lambda$ in (\ref{N}) from
the spectrum of large matrices by fitting its LSD to the surmised expression,
it inevitably involves a systematic error, $|\lambda/\Lambda-1| \lesssim 0.1$.
In a typical example of
extracting the pion decay constant from two-color QCD Dirac spectra \cite{nis},
one must instead use LSDs evaluated for infinite $N$
by the Nystr\"{o}m-type or analytic methods for fitting
in order to achieve accuracy better than $10\%$.


\begin{thebibliography}{99}

\bibitem{sbw}
S. Schierenberg, F. Bruckmann, and T. Wettig,
Phys. Rev. E {\bf 85}, 061130 (2012).

\bibitem{wig}
E. P. Wigner, in Conference on Neutron Physics by Time-of-Flight, Oak Ridge Natl. Lab. Report No. 2309, 1957 (unpublished).

\bibitem{lh}
G. Lenz and F. Haake, Phys. Rev. Lett. {\bf 67}, 1 (1991).

\bibitem{abps}
G. Akemann, E. Bittner, M.~J. Phillips, and L. Shifrin,
Phys. Rev. E {\bf 80}, 065201 (2009).

\bibitem{jmms}
M. Jimbo, T. Miwa, Y. M\^{o}ri, and M. Sato,
Physica D {\bf 1}, 80 (1980).

\bibitem{tw}
C.~A. Tracy and H. Widom,
Comm. Math. Phys. {\bf 177}, 727 (1996).

\bibitem{nis}
S.~M. Nishigaki, Prog. Theor. Phys. {\bf 128}, 1283 (2012);
Phys. Rev. D {\bf 86} (2012) in press [arXiv:1208.3452].

\bibitem{bor}
F. Bornemann, Math. Comp. {\bf 79}, 871 (2010).

\bibitem{mp}
M.~L. Mehta and A. Pandey, J. Phys. A: Math.~Gen. {\bf 16}, L601 (1983).

\bibitem{meh}
M.~L. Mehta, {\sl Random Matrices, 3rd ed.}, (Elsevier, New York, 2004), Sec. 14.

\bibitem{bbms}
J. Bloch, F. Bruckmann, N. Meyer, and S. Schierenberg, J. High Energy Phys. {\bf 08} (2012) 66.

\end{thebibliography}
\end{document}